\documentclass[prl,twocolumn,superscriptaddress,amssymb]{revtex4}
\usepackage{bm}
\usepackage{amsmath,amssymb,fancybox}
\usepackage{color}

\begin{document}
\title{Blind quantum computing can always
be made verifiable}
\author{Tomoyuki Morimae}
\email{tomoyuki.morimae@yukawa.kyoto-u.ac.jp}
\affiliation{Yukawa Institute for Theoretical Physics, Kyoto 
University, Kitashirakawa Oiwakecho, Sakyo-ku, Kyoto 606-8502, Japan}
\affiliation{JST, PRESTO, 4-1-8 Honcho, Kawaguchi, Saitama, 332-0012, Japan}

\begin{abstract}
Blind quantum computing enables a client, who does
not have enough quantum technologies, to delegate her quantum
computing to a remote quantum server in such a way that
her privacy is protected against the server. 
Some blind quantum computing protocols
can be made verifiable, which means that the client can
check the correctness of server's quantum computing.
Can any blind protocol always be made verifiable?
In this paper, we answer to the open problem affirmatively.
We propose a plug-in that makes any universal
blind quantum computing protocol
automatically verifiable. The idea is that the client blindly generates 
Feynman-Kitaev history states corresponding to the quantum 
circuit that solves client's problem and its complement circuit. 
The client can learn the solution of the problem and verify its correctness
at the same time by measuring energies of local Hamiltonians
on these states.
Measuring energies of local Hamiltonians
can be done with only single qubit measurements
of Pauli operators.
\end{abstract}
\maketitle  

Blind quantum computing~\cite{Childs,BFK,Aharonov,MF,topoblind,
CVblind,AKLTblind,Composability,distillation,
nogo,coherent,Lorenzo,Atul,Sueki,twostates,flow,overcoming} 
is a quantum cryptographic protocol
that enables a client (Alice), who does not have enough
quantum technologies,
to delegate her quantum computing to a remote quantum server, Bob,
in such a way that her privacy (the input, output, and program)
is protected against Bob.
(For a review, see Ref.~\cite{joe_review}).
The protocol proposed by Broadbent, Fitzsimons, and Kashefi (BFK)~\cite{BFK}
uses measurement-based quantum computing~\cite{MBQC}, and 
Alice needs only the ability of generating randomly-rotated single qubit 
states. The BFK protocol was experimentally realized with photonic
qubits~\cite{Barz_Science}.
The protocol proposed by Morimae and Fujii (MF)~\cite{MF},
which also uses measurement-based quantum computing, 
on the other hand,
requires Alice to do only
single-qubit measurements.

Several blind quantum computing protocols can be made
verifiable, which means that Alice can
check the correctness of Bob's quantum computing
in spite that her quantum technologies are severely limited~\cite{Aharonov,
FK,HM,topoveri,Takeuchi,inputverification,Michal,Andru1,Andru2,FH,Honda,
MTH,MT}.
(For a review, see Ref.~\cite{verification_review}.)
The verifiability is important in realistic cloud quantum computing, since
Bob might give Alice a wrong result deliberately, 
or it might be even the case that
what Bob actually has is not a real quantum computer
but a fake one.
Experimental realizations of verifiable
blind quantum computing were also done~\cite{Barz_NP,Chiara}.

To make blind quantum computing protocols
verifiable,
mainly two different types of techniques have been used.
The first type is so called the trap technique~\cite{Aharonov,FK}.
Alice hides some isolated qubits, which are called
``trap qubits", in the register, and later
checks that trap qubits are not disturbed by Bob.
Since Bob does not know the place of each trap qubit, he
will disturb a trap qubit with high probability if he deviates from
the correct procedure.
If the quantum computation is encoded with
a quantum error correcting code, 
the probability that Bob can change a logical state without
changing any trap qubit becomes exponentially small~\cite{FK}.
(If Bob wants to change the state of a logical qubit,
he has to change more than $d$ qubits, where $d$ is the code
distance. It increases the probability of he touching a trap qubit.)

The second type of the technique is called the stabilizer 
test~\cite{HM,MNS,MTH,MT}.
In the MF protocol, Bob generates graph states and sends
each qubit one by one to Alice.
Alice randomly choses some of graph states and
checks the correctness of them
by measuring stabilizer operators of the graph state.
Such tests can be done
with only single-qubit measurements of Pauli operators~\cite{HM,MNS,MTH,MT}.
It was shown in Refs.~\cite{HM,MNS,MTH,MT} 
that if Alice passes the stabilizer
test, a remaining state is close to the ideal graph state
on which she can do universal measurement-based quantum computing.

In this way, delegated quantum computing has two important properties,
the blindness and the verifiability.
Relations between them are still not clear.
For example, the blindness was believed to be 
necessary to achieve the verifiability,
but recently the belief has turned out to be wrong,
since a protocol that is verifiable but not necessarily blind
has been found~\cite{posthoc}.

Another open problem is
whether any blind protocol can always be made verifiable or not.
The BFK protocol~\cite{BFK} can be made verifiable 
(Fitzsimons-Kashefi (FK) protocol~\cite{FK}),
and the MF protocol~\cite{MF} can also be made verifiable 
(Hayashi-Morimae (HM) protocol~\cite{HM}).
However, the ways of making these blind protocols verifiable
are protocol specific, i.e., structures of the blind protocols
are exploited to make them verifiable. 
If someone finds a completely new blind protocol never seen before,
can we always make it verifiable?

In this paper, we solve the open problem
affirmatively.
We propose a ``plug-in" that makes any universal blind quantum
computing protocol automatically verifiable.
Our idea is based on the post hoc verification~\cite{posthoc}.
In the posthoc verification, the prover sends the verifier
the solution of a problem, and
the Feynman-Kitaev history state~\cite{KSV} corresponding to the quantum 
circuit that solves the problem or its complement circuit~\cite{posthoc}. 
The verifier can verify the correctness
of the solution by measuring the energy of a local Hamiltonian
on the history state.
Measuring energy of local Hamiltonians can 
be done with only single qubit measurements
of Pauli operators~\cite{MNS}.
Our idea is that Alice generates the two history states corresponding to the
solving circuit and the complement circuit by using the given universal
blind quantum computing protocol, and asks Bob to send them to Alice.
She can learn the solution and verify its correctness
at the same time by measuring their energies.
A more precise description of the procedure will be given later.
We will also see that our verification technique is different from 
and simpler than the two existing techniques
(i.e., the trap technique and the stabilizer technique)
explained above.

{\it Energy test}.---
Before explaining our procedure, we review 
the energy test~\cite{MNS},
which is an essential ingredient of our protocol.
Let 
\begin{eqnarray*}
H=\sum_Sd_SS
\end{eqnarray*}
be a Hamiltonian acting on $m$ qubits, 
where $d_S$ is a real number and 
$S$ is a tensor product of $m$ Pauli operators,
$X$, $Y$, $Z$, and $I$.

Let $\rho$ be an $m$-qubit state.
We call the following test ``the energy test for
$H$ on $\rho$":
\begin{itemize}
\item[1.]
Alice randomly chooses $S$ with probability 
$\frac{|d_S|}{\sum_S|d_S|}$~\cite{samplable}.
\item[2.]
Let 
\begin{eqnarray*}
S=\bigotimes_{j=1}^mP_j,
\end{eqnarray*}
where $P_j\in\{X,Y,Z,I\}$ for $j=1,2,...,m$.
If $P_j\neq I$,
Alice measures the $j$th qubit in the $P_j$-basis, 
and obtains the result $c_j\in\{\pm1\}$.
If $P_j=I$, Alice does nothing on the $j$th qubit,
and sets $c_j=+1$.
\item[3.]
If 
\begin{eqnarray*}
\prod_{j=1}^m c_j
=-\text{sign}(d_S), 
\end{eqnarray*}
Alice concludes that she passes the test.
\end{itemize}

The probability $p_{pass}$ that Alice passes the test
is
\begin{eqnarray*}
p_{pass}&=&\sum_S\frac{|d_S|}{\sum_S|d_S|}
{\rm Tr}
\Big(\frac{I^{\otimes m}-{\rm sign}(d_S)S}{2}\rho\Big)\\
&=&\frac{1}{2}-\frac{
\mbox{Tr}(H\rho)
}{\sum_S2|d_S|}.
\end{eqnarray*}

{\it Construction of a verifiable protocol}.---
Now we explain how to construct a verifiable blind protocol
from a blind protocol.
Let $L$ be a language in BQP.
Assume that Alice wants to know whether $x\in L$ or $x\notin L$ for
an instance $x$. 
Let $V_x$ be
the $n$-qubit quantum circuit corresponding to $x$,
which means that if $x\in L$, then
\begin{eqnarray*}
\big\|(|0\rangle\langle0|\otimes I^{\otimes n-1})V_x|0^n\rangle\big\|^2
\ge 1-2^{-r},
\end{eqnarray*}
and if $x\notin L$, then
\begin{eqnarray*}
\big\|(|0\rangle\langle0|\otimes I^{\otimes n-1})V_x|0^n\rangle\big\|^2
\le 2^{-r},
\end{eqnarray*}
where $r$ is any polynomial.
The circuit $V_x$ is written as
\begin{eqnarray*}
V_x= U_TU_{T-1}...U_1U_0,
\end{eqnarray*}
where $U_j$ ($j=0,1,2,...,T$) is a unitary gate acting on at most
a constant number of qubits, and $U_0=I^{\otimes n}$.
Each $U_j$ is taken from any standard universal gate set.
Let
$|\psi_0\rangle$ and $|\psi_1\rangle$ be
$m$-qubit Feynman-Kitaev history states corresponding
to $V_x$ and $(X\otimes I^{\otimes n-1})V_x$, respectively.
More precisely,
\begin{eqnarray*}
|\psi_0\rangle\equiv\frac{1}{\sqrt{T+1}}
\sum_{t=0}^T(U_t...U_0|0^n\rangle)\otimes|t\rangle
\end{eqnarray*}
and
\begin{eqnarray*}
|\psi_1\rangle&\equiv&\frac{1}{\sqrt{T+1}}
\Big[(X\otimes I^{\otimes n-1})(U_T...U_0|0^n\rangle)\otimes|T\rangle\\
&&+
\sum_{t=0}^{T-1}(U_t...U_0|0^n\rangle)\otimes|t\rangle\Big].
\end{eqnarray*}

It is known that there exist
$m$-qubit local Hamiltonians $H_0$ and $H_1$ such that:
\begin{itemize}
\item
If $x\in L$ then 
\begin{eqnarray*}
&&\langle\psi_0|H_0|\psi_0\rangle\le a,~\mbox{and}\\
&&{\rm Tr}(\sigma H_1)\ge b'~\mbox{for any $m$-qubit state $\sigma$}.
\end{eqnarray*}
\item
If $x\notin L$ then 
\begin{eqnarray*}
&&\langle\psi_1|H_1|\psi_1\rangle\le a',~\mbox{and}\\
&&{\rm Tr}(\sigma H_0)\ge b~\mbox{for any $m$-qubit state $\sigma$}.
\end{eqnarray*}
\end{itemize}
Here, $a,b,a'$ and $b'$ are certain parameters such that
$b-a\ge\frac{1}{\text{poly}(|x|)}$
and $b'-a'\ge\frac{1}{\text{poly}(|x|)}$.
It is easily shown by noticing the facts that 
BQP is in QMA, the local Hamiltonian problem is QMA-hard~\cite{KSV},
and BQP is closed under complement. 
(For details, see Ref.~\cite{posthoc}.
In Appendix, we also provide a detailed explanation
for the convenience of readers.)
Furthermore, it is known that
$H_0$ and $H_1$ can be two-local Hamiltonians with 
only $X$ and $Z$ operators~\cite{Love,CM}.

Let us define
\begin{eqnarray*}
\alpha&\equiv&\frac{1}{2}-\frac{a}{\sum_S2|d_S|},~~
\alpha'\equiv\frac{1}{2}-\frac{a'}{\sum_S2|d_S|},\\
\beta&\equiv&\frac{1}{2}-\frac{b}{\sum_S2|d_S|},~~
\beta'\equiv\frac{1}{2}-\frac{b'}{\sum_S2|d_S|}.
\end{eqnarray*}

Assume that a universal blind quantum computing protocol is given.
It can be the BFK protocol~\cite{BFK}, the MF protocol~\cite{MF}, 
or even a completely new protocol never seen before.
The following procedure makes the blind protocol
verifiable.
(We describe the procedure assuming that Bob is honest.
If Bob is malicious, the state of Eq.~(\ref{ideal}) is replaced with
any $(mk_0+mk_1)$-qubit state.)
\begin{itemize}
\item[1.]
By running the universal blind quantum computing protocol,
Alice blindly generates 
\begin{eqnarray}
|\psi_0\rangle^{\otimes k_0}\otimes|\psi_1\rangle^{\otimes k_1}
\label{ideal}
\end{eqnarray}
in Bob's place,
where $k_0$ and $k_1$ are some polynomials that will be specified
later~\cite{onetimepad}.
\item[2.]
Bob sends each qubit of Eq.~(\ref{ideal}) one by one to Alice.
\item[3.]
Alice does the energy test for $H_0$ 
on each $|\psi_0\rangle$.
Let $\eta_0$ be the number of times that she passes the test.
If
\begin{eqnarray*}
\frac{\eta_0}{k_0}\ge \frac{\alpha+\beta}{2},
\end{eqnarray*}
she outputs $\xi_0=1$.
Otherwise, she outputs $\xi_0=0$.
\item[4.]
Alice does the energy test for $H_1$ 
on each $|\psi_1\rangle$.
Let $\eta_1$ be the number of times that she passes the test.
If
\begin{eqnarray*}
\frac{\eta_1}{k_1}\ge\frac{\alpha'+\beta'}{2},
\end{eqnarray*}
she outputs $\xi_1=1$.
Otherwise, she outputs $\xi_1=0$.
\item[5.]
If $(\xi_0,\xi_1)=(1,0)$, Alice concludes $x\in L$.
If $(\xi_0,\xi_1)=(0,1)$, she concludes $x\notin L$.
Otherwise, she concludes that Bob is dishonest.
\end{itemize}
It is obvious that
this procedure does not degrade the blindness of the original protocol.
Therefore thus constructed verifiable blind protocol is
as secure as the original blind protocol.

{\it When Bob is honest}.---
First let us consider the case when Bob is honest.
What Alice receives is
the state of Eq.~(\ref{ideal}).

If $x\in L$,
the probability that Alice outputs $\xi_0=1$ is
\begin{eqnarray*}
{\rm Pr}\Big[\frac{\eta_0}{k_0}\ge \frac{\alpha+\beta}{2}\Big]
&=&1-
{\rm Pr}\Big[\frac{\eta_0}{k_0}< \frac{\alpha+\beta}{2}\Big]\\
&=&1-
{\rm Pr}\Big[\alpha-\frac{\eta_0}{k_0}>\frac{\alpha-\beta}{2}\Big]\\
&\ge&1-e^{-2k_0\frac{(\alpha-\beta)^2}{4}},
\end{eqnarray*}
and probability that Alice outputs $\xi_1=0$ is
\begin{eqnarray*}
{\rm Pr}\Big[\frac{\eta_1}{k_1}<\frac{\alpha'+\beta'}{2}\Big]
&=&
1-{\rm Pr}\Big[\frac{\eta_1}{k_1}\ge\frac{\alpha'+\beta'}{2}\Big]\\
&=&
1-
{\rm Pr}\Big[\frac{\eta_1}{k_1}-\beta'\ge\frac{\alpha'-\beta'}{2}\Big]\\
&\ge&
1-e^{-2k_1\frac{(\alpha'-\beta')^2}{4}}.
\end{eqnarray*}
Therefore, the probability that Alice concludes $x\in L$ is
larger than
\begin{eqnarray*}
\Big(1-e^{-2k_0\frac{(\alpha-\beta)^2}{4}}\Big)
\Big(1-e^{-2k_1\frac{(\alpha'-\beta')^2}{4}}\Big)
\ge(1-e^{-u})^2
\end{eqnarray*}
if we take $k_0$ and $k_1$ so that 
\begin{eqnarray*}
k_0&\ge&\frac{2u}{(\alpha-\beta)^2},\\
k_1&\ge&\frac{2u}{(\alpha'-\beta')^2}
\end{eqnarray*}
for any polynomial $u$.

In a similar way, we can show that if $x\notin L$,
the probability that Alice concludes $x\notin L$ is
larger than $(1-e^{-u})^2$ for any polynomial $u$.

{\it Bob is dishonest}.---
We next consider the case when Bob is dishonest.
What Alice receives is no longer the state of Eq.~(\ref{ideal}) but
any $(mk_0+mk_1)$-qubit state $\rho$.

Assume that $x\in L$. For any state $\rho$,
the probability that Alice outputs $\xi_1=1$ is
\begin{eqnarray*}
{\rm Pr}\Big[\frac{\eta_1}{k_1}\ge\frac{\alpha'+\beta'}{2}\Big]
&=&
{\rm Pr}\Big[\frac{\eta_1}{k_1}-\beta'\ge\frac{\alpha'-\beta'}{2}\Big]\\
&\le&
e^{-2k_1\frac{(\alpha'-\beta')^2}{4}}.
\end{eqnarray*}
Therefore, the probability that Alice concludes $x\notin L$ is
less than $e^{-u}$.

Assume that $x\notin L$. For any state $\rho$,
the probability that Alice outputs $\xi_0=1$ is
\begin{eqnarray*}
{\rm Pr}\Big[\frac{\eta_0}{k_0}\ge\frac{\alpha+\beta}{2}\Big]
&=&
{\rm Pr}\Big[\frac{\eta_0}{k_0}-\beta\ge\frac{\alpha-\beta}{2}\Big]\\
&\le&
e^{-2k_0\frac{(\alpha-\beta)^2}{4}}.
\end{eqnarray*}
Therefore, the probability that Alice concludes $x\in L$ is
less than $e^{-u}$.
Note that when Bob is dishonest, the results of Alice's energy
tests are not necessarily independent, since Bob might send
Alice any entangled state in stead of Eq.~(\ref{ideal}),
but using the standard argument
of the error reduction for QMA~\cite{aharonovQMA}, 
we can upperbound the soundness probability by considering
the case when
each energy test is an independent Bernoulli
trial with a success probability smaller than $\beta$ (or $\beta'$).

{\it Discussion}.---
In this paper, we have shown that 
any universal blind quantum computing protocol can always be made verifiable.
To conclude this paper, let us discuss the robustness of our 
verifiable protocol.
Even if Bob is honest, what Alice receives might be slightly
deviated from the state of Eq.~(\ref{ideal})
due to some imperfections of Bob's operations and noises
in the quantum channel from Bob to Alice.
However, as long as the deviated state is sufficiently
close to the ideal state in terms of the L1-norm, 
probabilities of passing energy tests
are not so much changed and 
therefore the $1/poly$ gap between the completeness
and the soundness should be maintained.
Furthermore, since universal blind quantum computing can be
done in the fault-tolerant way~\cite{topoblind,Takeuchi,FH}, 
and what Alice has to do in the energy test
is only $X$ and $Z$ measurements, which can be done transversally
in, for example, the CSS code, the full fault-tolerance
should be possible.
The detailed analysis with specific error parameters is, however,
beyond the scope of the present paper.

In this paper, we have considered only decision problems for simplicity,
but our result can be generalized to state generation tasks:
by measuring the energy, Alice can verify that Bob has honestly 
generated the correct Feynman-Kitaev history state.
From it, she can obtain $U_T...U_0|0^n\rangle$ with $1/poly$ probability
by measuring the clock register.

We also finally mention that
there are several verifiable blind quantum computing 
protocols that assume more than two servers who are entangling
but not communicating
with each other~\cite{RUV,Matt,Ji,Pan}.
These protocols are interesting because
Alice can be completely classical, 
but in this paper we have concentrated on the single-server
setup. It would be an interesting future research subject
to study relations between the blindness and the verifiability
in the multi-server setting.

TM thanks Yuki Takeuchi for his comments on the draft.
TM is supported by JST PRESTO and
the JSPS Grant-in-Aid
for Young Scientists (B) No.17K12637.

{\it Appendix}.---
Let $L$ be a language in BQP. It means that
for any polynomial $r$
there exists a uniformly-generated family $\{V_x\}_x$ of polynomial-size
quantum circuits such that
\begin{itemize}
\item
If $x\in L$ then 
$\big\|(|0\rangle\langle0|\otimes I^{\otimes n-1})V_x|0^n\rangle\big\|^2
\ge 1-2^{-r}$.
\item
If $x\notin L$ then 
$\big\|(|0\rangle\langle0|\otimes I^{\otimes n-1})V_x|0^n\rangle\big\|^2
\le 2^{-r}$.
\end{itemize}
Then $L$ is trivially in QMA with the verification circuit
$W_x\equiv V_x\otimes I^{\otimes w}$, and the yes witness state
$|0^w\rangle$, where $w$ is any polynomial.
Since the local Hamiltonian problem is QMA-hard,
there exists an local Hamiltonian $H_0$ acting on
$m$ qubits such that
\begin{itemize}
\item
If $x\in L$ then 
$\langle\psi_0|H_0|\psi_0\rangle\le a$
\item
If $x\notin L$ then 
for any $m$-qubit state $\sigma$,
${\rm Tr}(\sigma H_0)\ge b$.
\end{itemize}
Here $a$ and $b$ are certain parameters such that
$b-a\ge 1/poly(|x|)$.

Let us define $V_x'\equiv (X\otimes I^{\otimes n-1})V_x$.
Then,
\begin{itemize}
\item
If $x\in L$ then 
$\big\|(|0\rangle\langle0|\otimes I^{\otimes n-1})V_x'|0^n\rangle\big\|^2
\le 2^{-r}$.
\item
If $x\notin L$ then 
$\big\|(|0\rangle\langle0|\otimes I^{\otimes n-1})V_x'|0^n\rangle\big\|^2
\ge 1-2^{-r}$.
\end{itemize}
Therefore, in a similar argument, we can show that
there exists an local Hamiltonian $H_1$ acting on
$m$ qubits such that
\begin{itemize}
\item
If $x\notin L$ then 
$\langle\psi_1|H_1|\psi_1\rangle\le a'$
\item
If $x\in L$ then 
for any $m$-qubit state $\sigma$,
${\rm Tr}(\sigma H_1)\ge b'$.
\end{itemize}
Here $a'$ and $b'$ are certain parameters such that
$b'-a'\ge 1/poly(|x|)$.


\begin{thebibliography}{00}
\bibitem{samplable}
Note that we have assumed that 
the probability distribution $\{|d_S|/\sum_S|d_S|\}_S$
can be sampled in classical polynomial time.
Local Hamiltonians that we will consider later
satisfy such a property.
\bibitem{onetimepad}
In blind quantum computing, what is actually generated is not exactly
the state of Eq.~(\ref{ideal}), but the one up to a random Pauli
one-time pad~\cite{BFK,MF}. We can, however, correct 
the effect of the Pauli one-time pad by appropriately modifying
the basis of measurements that are done later on the state.
Therefore, without loss of generality, we can assume that the
state of Eq.~(\ref{ideal}) is generated.
\end{thebibliography}
\end{document}